\documentclass[letterpaper, 10 pt, conference]{ieeeconf}  % Comment this line out
\IEEEoverridecommandlockouts                              % This command is only
\usepackage{amsmath}
\usepackage{amssymb}
\usepackage{wasysym}
\usepackage{comment}

\usepackage{algorithm}
\usepackage{algpseudocode}

\newcommand{\always}{\mathsf{G}}
\newcommand{\eventually}{\mathsf{F}}
\newcommand{\Next}{\mathsf{X}}
\newcommand{\until}{~\mathsf{U}~}

\newcommand{\Traj}{\mathrm{Traj}}
\newcommand{\ITraj}{\mathrm{ITraj}}
\newcommand{\Inf}{\mathrm{Inf}}
\newcommand{\Fin}{\mathrm{Fin}}
\newcommand{\Viol}{\mathrm{Viol}}
\renewcommand{\Pr}{\mathrm{Pr}}

\DeclareMathOperator*{\argmin}{argmin}

\title{\LARGE \bf
Interpretable Apprenticeship Learning with \\ Temporal Logic Specifications
}

\author{Daniel Kasenberg and Matthias Scheutz% <-this % stops a space
\thanks{Authors are with the Department of Computer Science, Tufts University, Medford, MA 02155, USA.  The corresponding author is {\tt\small dmk@cs.tufts.edu}}%
}

\begin{document}

\maketitle
\thispagestyle{empty}
\pagestyle{empty}

%%%%%%%%%%%%%%%%%%%%%%%%%%%%%%%%%%%%%%%%%%%%%%%%%%%%%%%%%%%%%%%%%%%%%%%%%%%%%%%%
\begin{abstract}

	Recent work has addressed using formulas in linear temporal logic (LTL) as specifications for agents planning in Markov Decision Processes (MDPs).  We consider the inverse problem: inferring an LTL specification from demonstrated behavior trajectories in MDPs.  We formulate this as a multiobjective optimization problem, and describe state-based (``what actually happened'') and action-based (``what the agent expected to happen'') objective functions based on a notion of ``violation cost''.  We demonstrate the efficacy of the approach by employing genetic programming to solve this problem in two simple domains.

\end{abstract}

%%%%%%%%%%%%%%%%%%%%%%%%%%%%%%%%%%%%%%%%%%%%%%%%%%%%%%%%%%%%%%%%%%%%%%%%%%%%%%%%
\section{Introduction} \label{intro}

Apprenticeship learning, or learning behavior by observing expert demonstrations, allows artificial agents to learn to perform tasks without requiring the system designer to explicitly specify reward functions or objectives in advance.  Apprenticeship learning has been accomplished in agents in stochastic domains, such as Markov Decision Processes (MDPs), by means of \textit{inverse reinforcement learning} (IRL), in which agents infer some reward function presumed to underlie the observed behavior.  IRL has recently been criticized, especially in learning ethical behavior \cite{Arnold2017}, because the resulting reward functions (1) may not be easily explained, and (2) cannot represent complex temporal objectives.

Recent work (e.g., \cite{Ding2011,Fu2014,Wolff2012}) has proposed using linear temporal logic (LTL) as a specification language for agents in MDPs.  An agent in a stochastic domain may be provided a formula in LTL, which it must satisfy with maximal probability.  These approaches require the LTL specification to be specified a priori (e.g., by the system designer, although \cite{Dzifcak2009} construct specifications from natural language instruction).

This paper proposes combining the virtues of these approaches by inferring LTL formulas from observed behavior trajectories.  Specifically, this inference problem can be formulated as multiobjective optimization over the space of LTL formulas.  The two objective functions represent (1) the extent to which the given formula explains the observed behavior, and (2) the complexity of the given formula.  The resulting specifications are interpretable, and can be subsequently applied to new problems, but do not need to be specified in advance by the system designer.

The key contributions of this work are (1) the introduction of this problem and its formulation as an optimization problem; and (2) the notion of violation cost, and the state- and action-based objectives based on this notion.

In the remainder of the paper, we first discuss related work; we then describe our formulation of this problem as multiobjective optimization, defining a notion of ``violation cost'' and then describing state-based and action-based objectives, corresponding to inferring a specification from ``what actually happened'' and ``what the demonstrator expected to happen'' respectively.  We demonstrate the usefulness of the formulation by using genetic programming to optimize these objectives in two domains, called SlimChance and CleaningWorld.  We discuss issues pertaining to our approach and directions for future work, and summarize our results.

\section{Related Work}

The proposed problem draws primarily upon ideas from apprenticeship learning (particularly, inverse reinforcement learning), stochastic planning with temporal logic specifications, and inferring temporal logic descriptions of systems.

\subsection{Apprenticeship Learning}

Apprenticeship learning, the problem of learning correct behavior by observing the policies or behavioral trajectories of one or more experts, has predominantly been accomplished by \textit{inverse reinforcement learning} (IRL) \cite{Ng2000,Abbeel2004}.  IRL algorithms generally compute a reward function that ``explains'' the observed trajectories (typically, by maximally differentiating them from random behavior).  Complete discussion of the many types of IRL algorithms is beyond the scope of this paper.

The proposed approach bears some resemblance to IRL, particularly in its inputs (sets of finite behavioral trajectories).  Instead of computing a reward function based on the observed trajectories, however, the proposed approach computes a \textit{formula in linear temporal logic} that optimally ``explains'' the data.  This addresses the criticisms of \cite{Arnold2017}, who claim that IRL is insufficient in morally and socially important domains because (1) reward functions can be difficult for human instructors to understand and correct, and (2) some moral and social goals may be too temporally complex to be representable using reward functions.

\subsection{Stochastic Planning with Temporal Logic Specifications}

There has been a wealth of work in recent years on providing agents in stochastic domains (namely, Markov Decision Processes) with specifications in linear temporal logic (LTL).  The most straightforward approach is \cite{Ding2011}, which we describe further in section \ref{sec:ltlspec}.  The problem is to compute some policy which satisfies some LTL formula with maximal probability.

More sophisticated approaches consider the same problem in the face of uncertain transition dynamics \cite{Wolff2012,Fu2014}, partial observability \cite{Svorenova2015,Sharan2014}, and multi-agent domains \cite{Leahy2015,Guo2014}.  Also relevant to the proposed approach is the idea of ``weighted skipping'' that appears (in deterministic domains) in \cite{ReyesCastro2013,Tumova2013,Lahijanian2015}.

The problem of inferring LTL specifications from behavior trajectories is complementary to the problem of stochastic planning with LTL specifications, much as IRL is complementary to ``traditional'' reinforcement learning (RL).  Specifications learned using the proposed approach may be used for planning, and trajectories generated from planning agents may be used to infer the underlying LTL specification.

\subsection{Inferring Temporal Logic Rules from Agent Behavior}

The task of generating temporal logic rules that describe data is not a new one.  Automatic identification of temporal logic rules describing the behavior of software programs (in the category of ``specification mining'') has been attempted in, e.g., \cite{Gabel2008Symbolic,Gabel2010,lemieux2015}.  Lemieux et al's Texada \cite{lemieux2015} allows users to enter custom templates for formulas and retrieves all formulas satisfied by the observed traces up to user-defined support and confidence thresholds; this differs from the work of Gabel and Su, who decompose complex specifications into combinations of predefined templates.  Specifications in a temporal logic (rPSTL) have also been inferred from data in continuous control systems in \cite{Kong2014}.  Each approach deals with (deterministic) program traces.

The proposed approach is most strongly influenced by \cite{gecco2015}, which casts the task of inferring temporal logic specifications for finite state machines as a multiobjective optimization problem amenable to genetic programming.  Much of our approach follows from this work; our novel contribution is introducing the problem of applying such methods to agent behavior in stochastic domains, and in particular our notion of the \textit{violation cost} as an objective function.

\section{Preliminaries}
\label{prelim}

In this section we provide formal definitions of Markov Decision Processes (MDPs) and linear temporal logic (LTL); we then outline the approach taken in \cite{Ding2011} for planning to satisfy (with maximum probability) LTL formulas in MDPs.

\subsection{Markov Decision Processes}
\label{mdp}

The proposed approach pertains to agents in Markov Decision Processes (MDPs) augmented with a set, $\Pi$, of atomic propositions.  Since reward functions are not important to this problem, we omit them.  All notation and references to MDPs in this paper assume this construction.

Formally, a Markov Decision Process is a tuple \begin{equation*}\mathcal{M}=\langle S, U, A, P, s_0, \Pi, \mathcal{L} \rangle\end{equation*}
where
\begin{itemize}
	\item $S$ is a (finite) set of states;
	\item $U$ is a (finite) set of actions;
	\item $A:S \rightarrow 2^U$ specifies which actions are available in each state;
	\item $P:S\times U \times S \rightarrow [0,1]$ is a transition function, 
	with $P(s,a,s')=0$ if $a \notin A(s)$, so that $P(s,a,s')$ is the probability of transitioning to $s'$ by beginning in $s$ and taking action $a$;
	\item $s_0$ is an initial state;
	\item $\Pi$ is a set of atomic propositions; and
	\item $\mathcal{L}:S \rightarrow 2^\Pi$ is the labeling function, so that $\mathcal{L}(s)$ is the set of propositions that are true in state $s$. 
\end{itemize}

A \textit{trajectory} in an MDP specifies the path of an agent through the state space.  A \textit{finite trajectory} is a finite sequence of state-action pairs followed by a final state (e.g., $\tau = (s_0, a_0), \cdots, (s_{T-1}, a_{T-1}), s_T$); an \textit{infinite trajectory} takes $T \rightarrow \infty$, and is an infinite sequence of state-action pairs (e.g., $\tau=(s_0, a_0), (s_1, a_1), \cdots$).  A sequence (finite or infinite) is only a trajectory if $P(s_t, a_t, s_{t+1}) > 0$ for all $t \in \{0, \cdots, T-1\}$. We will denote by $\Traj_\mathcal{M}$ the set of all finite trajectories in an MDP $\mathcal{M}$, and by $\ITraj_\mathcal{M}$ the set of all infinite trajectories in $\mathcal{M}$. We will denote by $\tau|_T$ the $T$-time step truncation $(s_0, a_0), \cdots, (s_{T-1}, a_{T-1}), s_T$ of an infinite trajectory $r=(s_0, a_0), (s_1, a_1), \cdots$.

A \textit{policy} $M: \Traj_\mathcal{M} \times U \rightarrow [0,1]$ is a probability distribution over an agent's next action, given its previous (finite) trajectory.  A policy is said to be \textit{deterministic} if, for each trajectory, the returned distribution allots nonzero probability for only one action; we write $M:\Traj_\mathcal{M} \rightarrow U$.  A policy is said to be \textit{stationary} if the returned distribution depends only on the last state of the trajectory; we write $\pi:S \times U \rightarrow [0,1]$.

We denote $\ITraj^M_\mathcal{M}$ the set of all infinite trajectories that may occur under a given policy $M$.  More formally,
\begin{align*}
\ITraj^M_\mathcal{M} = \{\tau=(s_0, a_0), (s_1, a_1), \cdots \in \ITraj_\mathcal{M}: \\ M(\tau|_T, a_T) > 0 \textrm{ for all }T\}
\end{align*}

\subsection{Linear Temporal Logic}

Linear temporal logic (LTL) \cite{LTL} is a multimodal logic over propositions that linearly encodes time.  Its syntax is as follows:
\begin{align*}
\phi ::= &\top~|~\bot~|~p\mathrm{, where }~p\in\Pi~|~\neg \phi~|~\phi_1 \wedge \phi_2~|~\phi_1 \vee \phi_2~|\\& ~\phi_1 \rightarrow \phi_2~|~\Next \phi~|~\always \phi~|~\eventually \phi~|~\phi_1 \until \phi_2
\end{align*}
Here $\Next \phi$ means ``in the next time step, $\phi$''; $\always \phi$ means ``in all present and future time steps, $\phi$''; $\eventually \phi$ means ``in some present or future time step, $\phi$''; and $\phi_1 \until \phi_2$ means ``$\phi_1$ will be true until $\phi_2$ holds''.

The truth-value of an LTL formula is evaluated over an infinite sequence of \textit{valuations} $\sigma_0, \sigma_1, \cdots$, where for all $i$, $\sigma_i \subseteq \Pi$.  We say $\sigma_0, \sigma_1, \cdots \vDash \phi$ if $\phi$ is true given the infinite sequence of valuations $\sigma_0, \sigma_1, \cdots$.

There is thus a clear mapping between infinite trajectories and LTL formulas.  We abuse notation slightly and define \begin{equation*}\mathcal{L}((s_0,a_0),(s_1,a_1), \cdots)=\mathcal{L}(s_0), \mathcal{L}(s_1), \cdots\end{equation*}
We abuse notation further and say that for any $\tau \in \ITraj_\mathcal{M}$, $\tau \vDash \phi$ if $\mathcal{L}(\tau) \vDash \phi$.

We define the probability that a given policy satisfies an LTL formula $\phi$ by
\begin{equation*}
\Pr^M_\mathcal{M}(\phi) = \Pr\{\tau \in \ITraj^M_\mathcal{M}:\tau \vDash \phi\}
\end{equation*}
That is, the probability that an infinite trajectory under $M$ will satisfy $\phi$.

Each LTL formula can be translated into a \textit{deterministic Rabin automaton} (DRA), a finite automaton over infinite words.  DRAs are the standard approach to model checking for LTL.  A DRA is a tuple 
\begin{equation*}
\mathcal{D} = \langle Q, \Sigma, \delta, q_0, F \rangle
\end{equation*}
where
\begin{itemize}
	\item $Q$ is a finite set of states;
	\item $\Sigma$ is an alphabet (in this case, $\Sigma = 2^\Pi$, so words are infinite sequences of valuations);
	\item $\delta:Q \times \Sigma \rightarrow Q$ is a (deterministic) transition function;
	\item $q_0$ is an initial state; and
	\item $F = \{ (\Fin_1, \Inf_1), \cdots, (\Fin_k,\Inf_k) \}$, where $\Fin \subseteq Q$, $\Inf \subseteq Q$ for all $(\Inf, \Fin) \in F$ specifies the acceptance conditions.
\end{itemize}

% specify \tau for trajectories; r for DRA runs.

A \textit{run} $r = q_0, q_1, \cdots$ of a DRA is an infinite sequence of DRA states such that there is some word $\sigma_0 \sigma_1 \cdots$ such that $\delta(q_i, \sigma_i) = q_{i+1}$ for all $i$.  A run $r$ is considered accepting if there exists some $(\Fin, \Inf) \in F$ such that for all $q \in \Fin$, $q$ is visited only finitely often in $r$, and $\Inf$ is visited infinitely often in $r$.

\subsection{Stochastic Planning with LTL Specifications}
\label{sec:ltlspec}
Planning to satisfy a given LTL formula $\phi$ within an MDP $\mathcal{M}$ with maximum probability generally follows the approach of \cite{Ding2011}.

The planning agent runs the DRA for $\phi$ alongside $\mathcal{M}$ by constructing a \textit{product MDP} $\mathcal{M}^\times$ which augments the state space to include information about the current DRA state.

Formally, the product of an MDP $\mathcal{M} = \langle S, U, A, T, s_0, \Pi, \mathcal{L} \rangle$ and a DRA $\mathcal{D} = \langle Q, 2^\Pi, \delta, q_0, F \rangle$ is an MDP
\begin{equation*}
\mathcal{M}^\times = \langle S^\times, U^\times, A^\times, P^\times, s^\times_0, \Pi^\times, \mathcal{L}^\times \rangle
\end{equation*}
where
\begin{itemize}
	\item $S^\times = S \times Q$;
	\item $U^\times = U; A^\times = A$;
	\item $P^\times((s,q), a, (s', q')) =$
	\begin{equation*}\begin{cases} P(s,a,s') &\textrm{if }q'=\delta(q, \mathcal{L}(s'))\\ 0&\textrm{otherwise} \end{cases}\end{equation*}
	\item $s^\times_0 = (s_0, \delta(q_0, \mathcal{L}(s_0)))$
	\item $\Pi^\times =\Pi;$ and $\mathcal{L}^\times = \mathcal{L}$.
\end{itemize}
The agent constructs the product MDP $\mathcal{M}^\times$, and then computes its \textit{accepting maximal end components} (AMECs).  An \textit{end component} $\mathcal{E}$ of an MDP $\mathcal{M}^\times$ is a set of states $S_\mathcal{E} \subset S^\times$ and an action restriction (mapping from states to sets of actions) $A_\mathcal{E}:S_\mathcal{E} \rightarrow 2^U$ such that (1) any agent in $S_\mathcal{E}$ that performs only actions as specified by $A_\mathcal{E}$ will remain in $S_\mathcal{E}$; and (2) any agent with a policy assigning nonzero probability to all actions in $A_\mathcal{E}$ is guaranteed to eventually visit each state in $A_\mathcal{E}$ infinitely often.  

An end component thus specifies a set of states $S_\mathcal{E}$ such that with an appropriate choice in policy, the agent can guarantee that it will remain in $S_\mathcal{E}$ forever, and that it will reach every state in $S_\mathcal{E}$ infinitely often.  An end component is maximal if it is not a proper subset of another end component.  An end component is \textit{accepting} if there is some $(\Fin, \Inf) \in F$ such that (1) if $q \in \Fin$, then $(s,q) \notin S_\mathcal{E}$ for all $s \in S$; and (2) there exists some $q \in \Inf, s \in S$ such that $(s, q) \in S_\mathcal{E}$.  In this case, by entering $S_\mathcal{E}$ and choosing an appropriate policy (for instance, a uniformly random policy over $A_\mathcal{E}$), the agent guarantees that the DRA run will be accepting.  A method for computing the AMECs of the product MDP is found in \cite{Baier2008}.

The problem of satisfying $\phi$ with maximal probability is thus reduced to the problem of reaching, with maximal probability, any state in any AMEC.  \cite{Ding2011} shows how this can be solved using linear programming.

\section{Optimization Problem}

Suppose that an agent is given some set of finite behavior trajectories $\tau^1, \cdots, \tau^m \in \Traj_\mathcal{M}$, where $\tau^i = (s^i_0, a^i_0), \cdots, (s^i_{T_i-1}, a^i_{T_i-1}), s^i_{T_i}$ for all $i\in \{1, \cdots, m\}$.

We refer to the agent whose trajectories are observed as the \textit{demonstrator}, and the agent that observes the trajectories as the \textit{apprentice}.  There may be several demonstrators satisfying the same objectives; this does not affect the proposed approach.

The proposed problem is to infer an LTL specification that well (and succinctly) explains the observed trajectories.  This can be cast as a multiobjective optimization problem with two objective functions:

\begin{enumerate}
	\item An objective function representing how well a candidate LTL formula explains the observed trajectories (and distinguishes them from random behavior); and
	\item An objective function representing the complexity of a candidate LTL formula.
\end{enumerate}

This section proceeds by describing a notion of ``violation cost'' (and defining the violation cost of infinite trajectories and policies) and using it to define two alternate objective functions representing (a) how well a candidate formula explains the \textit{actual observed state sequence} (a ``state-based'' objective function), and (b) how well a candidate formula explains the \textit{actions} of the demonstrator in each state (an ``action-based'' objective function).  We then describe the simple notion of formula complexity we will utilize, and formulate the optimization problem.

\subsection{Violation Cost}

We are interested in computing LTL formulas that well explain the demonstrator's trajectories.  These formulas should be satisfied by the observed behavior, but not by random behavior within the same MDP (since, for example, the trivial formula $\always~\top$ will be satisfied by the observed behavior, but also by random behavior).  Ideally we could assign a ``cost'' either to trajectories (finite or infinite) or to policies (and, particularly, to the uniformly random policy in $\mathcal{M}$), where the cost of a trajectory or policy corresponds to its adherence to or deviance from the specification.  Given such a cost function $C$, the objective would be to minimize $\sum_i \left( C(\tau^i) - C(\pi_{rand})\right)$, where $\pi_{rand}:S\times U \rightarrow [0,1]$ is the uniformly-random (stationary) policy over $\mathcal{M}$:
\begin{equation*}
\pi_{rand}(s,a) = \begin{cases} \frac{1}{|A(s)|} &\textrm{if }a \in A(s) \\ 0 & \textrm{otherwise}\end{cases}
\end{equation*}

The obvious choice of such a cost function (over infinite trajectories $\tau$) would be the indicator function $\mathbf{1}_{\tau \nvDash \phi}$ which returns $0$ if $\tau \vDash \phi$ and $1$ otherwise; this function may be extended to general policies $M$ by $1-\Pr^M_\mathcal{M}(\phi)$.  This function, however, cannot distinguish between small and large deviances from the specification.  For example, given the specification $\always~p$, this function cannot differentiate between $\tau$ such that $p$ is \textit{almost always} true and $\tau$ such that $p$ is \textit{never} true.  We thus propose a more sophisticated cost function.

For $\tau \in \ITraj_\mathcal{M}$, $N$ a set of nonnegative integers, we define $\tau \backslash N$ to be the subsequence of $\tau$ omitting the state-action pairs with time step indices in $N$.  For example, $(s_0, a_0), (s_1, a_1), (s_2, a_2), , \cdots \backslash \{1\} = (s_0, a_0), (s_2, a_2), \cdots$.  Each time step with an index in $N$ is said to be ``skipped''.

We define the \textit{violation cost} of an infinite trajectory $\tau \in \ITraj_\mathcal{M}$ subject to the formula $\phi$ as the (discounted) minimum number of time steps that must be skipped in order for the agent to satisfy the 
formula:

\begin{equation}
\Viol_\phi(\tau) = \min\limits_{\substack{N \subseteq \mathbb{N}_0 \\ \tau \backslash N \vDash \phi}} \sum\limits_{t=0}^\infty \gamma^t \mathbf{1}_{t \in N}
\end{equation}
Note that if $\tau \vDash \phi$, then $\Viol_\phi(\tau) = 0$.

% BEGIN bad section
In order to define a similar measure for policies, we must construct an augmented product MDP $\mathcal{M}^\otimes$, which is similar to $\mathcal{M}^\times$ as described in section \ref{sec:ltlspec}, but allows an agent to ``skip'' states by performing at each time step (simultaneously with their normal actions), a ``DRA action'' $\tilde{a} \in \{keep, susp\}$, where $keep$ causes the DRA to transition as usual, and $susp$ causes the DRA to not update in response to the new state.

Formally, given an MDP $\mathcal{M} = \langle S, U, A, T, s_0, \Pi, \mathcal{L} \rangle$ and a DRA $\mathcal{D} = \langle Q, \Sigma, \delta, q_0, F \rangle$ corresponding to the specification $\phi$, we may construct a product MDP $\mathcal{M}^\otimes = \langle S^\otimes, U^\otimes, A^\otimes, T^\otimes, s^\otimes_{-1}, \Pi^\otimes, \mathcal{L}^\otimes \rangle$ as follows:
\begin{itemize}
	\item $S^\otimes = \left(S \cup \{s_{-1}\}\right) \times Q$
	\item $U^\otimes = (U \cup \{a_{-1}\}) \times \tilde{U}$, where $\tilde{U} = \{ keep, susp \}$
	\item $A^\otimes((s,q)) = \begin{cases}\{a_{-1}\} \times \tilde{U}&\textrm{if }s=s_{-1} \\A(s) \times \tilde{U}&\textrm{otherwise}\end{cases}$
	\item $s^\otimes_{-1} = (s_{-1}, q_0)$
	\item $P^\otimes(s^\otimes_{-1}, (a_{-1}, keep), (s_0, \delta(q_0, \mathcal{L}(s_0)))) = 1$
	\item $P^\otimes(s^\otimes_{-1}, (a_{-1}, susp), (s_0, q_0)) = 1$
	\item Otherwise, $P^\otimes((s,q), (a, \tilde{a}), (s', q')) =$
	\begin{equation*}\begin{cases} P(s,a,s') & \textrm{if }q'=\delta(q, \mathcal{L}(s'))\textrm{ and }\tilde{a}=keep\\ P(s,a,s') & \textrm{if }q'=q\textrm{ and }\tilde{a}=susp\\ 0 & \textrm{otherwise} \end{cases}
	\end{equation*}
	\item $\Pi^\otimes = \Pi, \mathcal{L}^\otimes = \mathcal{L}$
\end{itemize}
The state $s_{-1}$ and action $a_{-1}$ are added so that the agent may choose to ``skip'' time step $t=0$.  This is necessary for the case that $s_0$ violates the specification.

Note that the transition dynamics of $\mathcal{M}^\otimes$ are such that $N$ (the set of ``skipped'' time step indices) can be defined as

\begin{equation}
N = \{ t \in \mathbb{N}_0 : \tilde{a}_{t-1} = susp \}
\end{equation}

Define the \textit{transition cost} $s^\otimes, (a, \tilde{a}), {s^\otimes}'$ in $\mathcal{M}^\otimes$ as \begin{equation}TC(s^\otimes, (a, \tilde{a}), {s^\otimes}') = \mathbf{1}_{\tilde{a}=susp}\end{equation}

The violation cost of a (non-product) trajectory $\tau$ can then be rewritten as a discounted sum of the transition costs at each stage, minimized over the DRA actions $\tilde{a}_{-1}, \tilde{a}_0, \tilde{a}_1, \cdots$, subject to the constraint that the DRA run from carrying out $\tau$ and the DRA actions must be accepting.  This indicates that the violation cost of a policy $\pi$ may be thought of as the state-value function for the policy $\pi$ with respect to $TC$.  Indeed, we will define the violation cost of a policy this way.

We define a \textit{product policy} to be a stationary policy $\pi^\otimes:S^\otimes \times (U \cup \{a_{-1}\}) \rightarrow [0,1]$.  When we consider the violation cost of a policy, we will assume a product policy of this form.

There are two reasons for this.  First, when evaluating a candidate specification, we wish to assume the demonstrator had knowledge of that specification (or else we would be unable to notice complex temporal patterns in agent behavior), and thus that the demonstrator's policy is over product states.  Second, we wish to allow the demonstrator to observe the new (non-product) state $s_t$ before deciding whether to ``skip'' time step $t$.  That is, $s_t$ should be observed before $\tilde{a}_{t-1}$ is chosen, which is inconsistent with the typical policy $\pi:S^\otimes \times U^\otimes \rightarrow [0,1]$ over the product space.

We can easily construct a product policy from the uniformly random policy on $\mathcal{M}$.  We define $\pi^\otimes_{rand}((s,q), a) = \pi_{rand}(s,a)$
for all $s \in S, a \in A$. 

Upon constructing the product MDP $\mathcal{M}^\otimes$, we compute its AMECs (as in section \ref{sec:ltlspec}).  Then let $S_{good} = \bigcup\limits_{i \in \{1, \cdots, p\}}S_{\mathcal{E}_i}$, and let $S_{bad}$ be the set of states in the product space from which no state in $S_{good}$  can be reached; these can be determined by breadth-first search.

We can use a form of the Bellman update equation to perform policy evaluation on a product policy $\pi^\otimes$.  For each state $s^\otimes \in S_{bad}$, we initialize the cost of this state to the maximum discounted cost, $\frac{1}{1-\gamma}$, and we do not update these costs.  This is done to enforce the constraint that the minimization should be over accepting DRA runs.  Otherwise, the violation cost will always be trivially zero (since $\tilde{a}=keep$ will always be picked).  The update equation has the following form:
\begin{align}
\Viol^{(k+1)}((s,q)) \gets \Big(\sum\limits_{a \in A(s)}\pi^\otimes((s,q),a)\Big(\sum\limits_{s' \in S}P(s,a,s') \nonumber \\ \min\{ 1 + \gamma \Viol^{(k)}(s',q), \gamma \Viol^{(k)}(s',\delta(q,\mathcal{L}(s')))  \}\Big)\Big) \label{belupdate}
\end{align}

The $\min\{ \cdot \}$ in (\ref{belupdate}) is where the optimization over $\tilde{a}$ (implicitly) occurs.  Choosing $\tilde{a} =susp$ incurs a transition cost of $1$ and then causes the DRA to remain in state $q$; choosing $\tilde{a}= keep$ incurs no transition cost, but causes the DRA to transition to state $\delta(q, \mathcal{L}(s'))$.  The ability for the demonstrator to optimize over $\tilde{a}$ after observing the new state $s'$ corresponds to the location of the $\min\{ \cdot \}$ in the Bellman update.

We define the violation cost of a policy as the function that results when running this update equation to convergence:

\begin{equation}
\Viol^{\pi^\otimes}_\phi((s,q)) = \lim_{k \rightarrow \infty}\Viol^{(k)}((s,q))
\end{equation}

We now consider state-based (``what actually happened'') and action-based (``what the agent expected to happen'') objective functions, for explaining sets of finite trajectories.

The crux of both the state- and action-based objective functions is Algorithm \ref{algqseq}.  Given a finite sequence of states $s_0, \cdots, s_T$, Algorithm \ref{algqseq} determines the ``optimal product-space interpretation'' of $s_0, \cdots, s_T$.  We define a \textit{product space interpretation} of a sequence of states $s_0, \cdots, s_T$ in an MDP $\mathcal{M}$ as a sequence of DRA states $q_0, \cdots, q_{T+1}$ such that, for all $i \in \{ 1, \cdots, T+1 \}$, either $q_i = q_{i-1}$, or $q_i = \delta(q_{i-1}, \mathcal{L}(s_{i-1}))$.  That is, a product-space interpretation specifies a possible trajectory in $\mathcal{M}^\otimes$ that is consistent with the observed trajectory in $\mathcal{M}$.

\begin{algorithm}[t]
	\caption{Best DRA state sequence for finite state sequence $s_0, \cdots, s_T$}\label{algqseq}
	\begin{algorithmic}[1]
		\small
		\Function{GetRabinStateSequence}{$Viol^{\pi^\otimes_{rand}}_\phi$, $\mathcal{M}^\otimes$, $S_{bad}$, $s_0, \cdots, s_T$}
		\State{$C_t[s^\otimes] \gets \infty$ for all $t \in \{-1,0,\cdots, T \}, s^\otimes \in S^\otimes$}\label{alg1:Cinit}
		\State{$R_{-1} = \{s^\otimes_{-1}\}$}\label{alg1:Rminus1}
		\State{$C_{-1}[s^\otimes_{-1}]\gets 0$}\label{alg1:Cminus1}
		\State{$seq_{-1}[s^\otimes_{-1}] \gets q_0$}\label{alg1:seqminus1}
		\For{$t\in\{0, \cdots, T\}$}
		\State{$R_t = \emptyset$}\label{alg1:Rtnull}
		\For{$(s,q) \in R_{t-1}$}
		\State{$q' \gets \delta(q, \mathcal{L}(s_t))$}
		\State{$R_t \gets R_t \cup \{ (s_t, q), (s_t,q') \} $}\label{alg1:Rt}
		\If{$C_{t-1}[(s,q)] + \gamma^t < C_t[(s_t,q)]$}
		\State{$C_t[(s_t, q)] \gets C_{t-1}[(s,q)] + \gamma^t$}\label{alg1:Csusp}
		\State{$seq_t[(s_t, q)] \gets (seq_{t-1}[(s,q)],q)$}\label{alg1:seqsusp}
		\EndIf
		\If{$C_{t-1}[(s, q)] < C_{t-1}[(s_t, q')]$}
		\State{$C_t[(s_t,q')] \gets C_{t-1}[(s,q)] $}\label{alg1:Ckeep}
		\State{$seq_t[(s_t, q')] \gets (seq_{t-1}[(s,q)], q')$}\label{alg1:seqkeep}
		\EndIf
		\EndFor
		\EndFor
		\State{$s^\otimes_T \gets \argmin\limits_{s^\otimes \in R_T \backslash S_{bad}}C_T[s^\otimes] + \gamma^{T+1} \Viol^{\pi^\otimes_{rand}}_\phi(s^\otimes)$}
		\Return {$seq_T[s^\otimes_T]$, $C_T[s^\otimes_T] + \gamma^{T+1} \Viol^{\pi^\otimes_{rand}}_\phi(s^\otimes_T)$}
		\EndFunction
	\end{algorithmic}
\end{algorithm}

Algorithm \ref{algqseq} uses dynamic programming to determine, for each time step $t$, the set of states $R_t$ of DRA states that the demonstrator could be in at time $t$ (lines \ref{alg1:Rminus1},\ref{alg1:Rtnull}, and \ref{alg1:Rt}), as well as the minimal violation cost $C_t[q_{t+1}]$ that would need to be accrued in order to be in each such state $q_{t+1}$ (lines \ref{alg1:Cinit},\ref{alg1:Cminus1}, \ref{alg1:Csusp}, and \ref{alg1:Ckeep}).  The sequence $seq_t[(s_t, q_{t+1})]=q_0, \cdots, q_{T+1}$ that achieves this minimal cost is also computed (lines \ref{alg1:seqminus1}, \ref{alg1:seqsusp}, and \ref{alg1:seqkeep}).

The apprentice then assumes that the demonstrator acted randomly from time step $T+1$ onward.  Although this assumption is probably incorrect, it is not entirely unreasonable, since it avoids the assumption that the demonstrator attempted to satisfy the formula after time step $T+1$, which would artificially drive the net violation cost down; this allows the apprentice to reuse values that are already computed in order to evaluate the random policy.

Employing this assumption, the apprentice determines the optimal product-space interpretation as $seq_T[s^\otimes_T]$, where
\begin{equation}
s^\otimes_T =\argmin\limits_{s^\otimes \in R_T \backslash S_{bad}}C_T[s^\otimes] + \gamma^{T+1} \Viol^{\pi^\otimes_{rand}}_\phi(s^\otimes)\label{stimest}
\end{equation}

\subsubsection{State-based objective function}

We first consider an approach to estimating the violation cost of a finite trajectory that considers only the states visited in the trajectory, ignoring the demonstrator's actions.

The state-based violation cost is the minimand of (\ref{stimest}), which is the second value returned by Algorithm \ref{algqseq}:
\begin{equation}
\Viol^S_\phi(\tau) = C_T[s^\otimes_T] + \gamma^{T+1} \Viol^{\pi^\otimes_{rand}}_\phi(s^\otimes_T)
\end{equation}
Thus the state-based objective function for $\tau^1,\cdots, \tau^m$ is the sum of the estimated violation costs of all observed finite trajectories, less $m$ times the expected violation cost of the random policy from the initial state:
\begin{equation}
\mathrm{Obj}^S(\phi) = \left(\sum\limits_{i=1}^m\textrm{Viol}^S_\phi(\tau^i)\right) - m\Viol^{\pi^\otimes_{rand}}_\phi(s^\otimes_{-1})\label{objS}
\end{equation}
The main drawback of the state-based approach is that by ignoring the observed actions, the apprentice neglects a crucial detail: that what the demonstrator ``\textit{expected}'' or ``\textit{intended}'' to satisfy may differ from what actually \textit{was} satisfied.  The fact that $p$ did not occur does not mean that the demonstrator was not attempting to make $p$ occur with maximal probability, particularly if $p$ is a very rare event.  To solve this problem, we consider an action-based approach.

\subsubsection{Action-based objective function}

We now consider estimating the violation cost of a finite trajectory $\tau$ by using the observed state-action pairs to compute a partial policy over the product MDP $\mathcal{M}^\otimes$.

\begin{algorithm}[t]
	\caption{Action-based violation cost of set of finite set of finite state-action trajectories $\tau^1, \cdots, \tau^m$ where $\tau^i= (s^i_0, a^i_0), \cdots, (s^i_{T_i-1}, a^i_{T_i-1}), s^i_{T_i}$}\label{algab}
	\begin{algorithmic}[1]
		\small
		\Function{ActionBasedViolationCost}{$Viol^{\pi^{rand}_\mathcal{M}}_\phi$, $\mathcal{M}^\otimes$, $S_{bad}$, $\tau^1, \cdots, \tau^m$}
		\State $A^*[s^\otimes] \gets \emptyset$ for $s^\otimes \in S^\otimes$
		\For{$i \in \{1, \cdots, m\}$}
		\State $q^i_0, \cdots, q^i_{T_i+1} ,V \gets$ \Call{GetRabinStateSequence}{$s^i_0, \cdots, s^i_{T_i}$}\label{algab:call}
		\For{$t \in \{-1,0, \cdots, T_i-1\}$}
		\State{$A^*[(s^i_t, q^i_{t+1})] \gets A^*[(s^i_t, q^i_{t+1})] \cup \{ a^i_t \} $}\label{algab:astar}
		\EndFor
		\EndFor
		\For{$s^\otimes = (s, q) \in S^\otimes$}
		\If{$A^*[s^\otimes] = \emptyset$}
		\State{$A^*[s^\otimes] = A(s)$}\label{algab:astardefault}
		\EndIf
		\EndFor
		\State{Compute $\Viol^{\pi^\otimes_{A^*}}_\phi$ using (\ref{belupdate})} \label{algab:belupdate}
		\\ \Return{$\Viol^{\pi^\otimes_{A^*}}_\phi(s^\otimes_{-1})$}
		\EndFunction
	\end{algorithmic}	
\end{algorithm}

To compute the action-based violation cost of a set of trajectories $\tau^1, \cdots, \tau^m$ (Algorithm \ref{algab}), the apprentice first runs Algorithm \ref{algqseq} to determine the optimal product-space interpretation $q^i_0, \cdots, q^i_{T+1}$ for each trajectory $\tau^i$ (line \ref{algab:call}), and uses this to compute the resulting product-space sequence $s^{\otimes i}_{-1}, \cdots, s^{\otimes i}_T$ where $s^{\otimes i}_t = (s^i_t, q^i_{t+1})$.

The assumption that for each $i \in \{1, \cdots, m\}$, $t \in \{0, \cdots, T_i -1\}$, the demonstrator performed $a^i_t$ when in the inferred product MDP state $s^{\otimes i}_t$, induces an action restriction (lines \ref{algab:astar} and \ref{algab:astardefault}) $A^*:S^\otimes \rightarrow 2^U$ where
\begin{equation*}
A^*(s^\otimes) = \begin{cases}
\bigcup\limits_{\substack{i, t:\\ s^\otimes = s^{\otimes i}_t}} \{a^i_t\} & \textrm{if }\bigcup\limits_{\substack{i, t:\\ s^\otimes = s^{\otimes i}_t}} \{a^i_t\} \neq \emptyset \\ A^\otimes(s^\otimes) & \textrm{otherwise}
\end{cases}
\end{equation*}
The apprentice may then compute, using the Bellman update (\ref{belupdate}), the violation cost of the policy $\pi^{rand}_{A^*}(s^\otimes)$ that uniformly-randomly chooses an action from $A^*(s^\otimes)$ at each state $s^\otimes$ (line \ref{algab:belupdate}):
\begin{equation*}
\pi^\otimes_{A^*}(s^\otimes, a)= \begin{cases}\frac{1}{|A^*(s^\otimes)|} & \textrm{if }a \in A^*(s^\otimes) \\ 0 &\textrm{otherwise}\end{cases}
\end{equation*}
The action-based objective function is then
\begin{equation}
\mathrm{Obj}^A(\phi) = \Viol^{\pi^\otimes_{A^*}}_\phi(s^\otimes_{-1}) - \Viol^{\pi^\otimes_{rand}}_\phi(s^\otimes_{-1})\label{objA}
\end{equation}
\subsection{Formula Complexity}
\label{fc}
Given two formulas that equally distinguish between the observed behavior and random behavior, we wish to select the less complex of the two.  Here it suffices to simply minimize the number of nodes in the parse tree for the LTL formula (that is, the total number of symbols in the formula).  There are also more sophisticated ways to evaluate formula complexity (such as that used in \cite{gecco2015}), but they are not necessary for our purposes.

\subsection{Multiobjective Optimization Problem}

Given some set of finite trajectories $\tau^1, \cdots, \tau^m$, we thus frame the problem of inferring some LTL formula $\phi$ that describes $\tau^1, \cdots, \tau^m$ as
\begin{equation*}
\min\limits_{\phi \in \mathrm{LTL}}(\mathrm{Obj}(\phi), FC(\phi))
\end{equation*}
where $\mathrm{Obj}$ is either $\mathrm{Obj}^S$, as described in (\ref{objS}), or $\mathrm{Obj}^A$, as described in (\ref{objA}); $FC$ is formula complexity (in this case, the number of nodes in the formula) as specified in section \ref{fc}.
\section{Examples}

To demonstrate the effectiveness of the proposed objective functions, we employed genetic programming to evolve a set of LTL formulas (where formulas are represented by their parse trees) in two domains.  A summary of the domains used is in Table \ref{tbl:runtime}.  In all demonstrations, we used MOEAFramework \cite{hadka2012} for genetic programming, using standard tree crossover and mutation operations \cite{koza1992}.  We consider (separately) the state-based and action-based objectives. NSGA-II over each set of objectives was run for $50$ generations with a population size of $100$.  This process was repeated $20$ times.  We employed BURLAP \cite{burlap} for MDP planning, and Rabinizer 3 \cite{rabinizer3} for converting LTL formulas to DRAs.   In each case, we restricted search to formulas of the form $\always~\phi$.

\begin{table*}
	\begin{center}
		\caption{Summary of example domains, with run times for state-based/action-based objectives}\label{tbl:runtime}
		\begin{tabular}{c|c|c|c|c|c}
			Domain & \# States & \# Actions & ``Actual'' specification & Time, state-based (s) & Time, action-based (s)\\ 
			\hline \hline
			SlimChance & 2 & 2 & $\always~good$ & $174.8 \pm 18.0$ & $372.0 \pm 43.3$ \\
			CleaningWorld & 77 & 5 & $\always~(\Next(vacuum) \until roomClean)$ & $19139.3 \pm 671.2$ & $32932.3 \pm 1755.0$
		\end{tabular}
	\end{center}
	\vspace{-2em}
\end{table*}

The tables in this section show formulas that are \textit{Pareto efficient} in at least two NSGA-II runs - that is, there were no solutions within those runs that outperformed them on both objectives.  For any Pareto inefficient formula $\phi$, there is some formula $\phi'$ which both (1) better explains the demonstrated trajectories (as measured by the violation-cost objective function) and (2) is simpler.  Thus it is reasonable to restrict consideration to only Pareto efficient solutions.

\subsection{SlimChance domain}

The SlimChance domain consists of two states: $s_{GOOD}$, a ``good'' state, and $s_{BAD}$, a ``bad'' state.  The agent has two actions: $try$, and $notry$.  If the agent performs $notry$, the next state is always $s_{BAD}$; if the agent performs $try$, the next state is $s_{GOOD}$ with small probability $\epsilon= 0.01$ and $s_{BAD}$ otherwise.  Thus, performing the $try$ action is ``trying'' to make the good state occur, but will rarely succeed.

The set $\Pi$ of atomic propositions for this problem consists of a single proposition $good$, which is true in $s_{GOOD}$ but false in $s_{BAD}$.  We then suppose that the agent is attempting to satisfy the simple LTL formula $\always~good$.

A demonstrator attempting to minimize violation cost generated three trajectories of 10 time steps each.  This resulted in the following trajectories (note that $\tau^1=\tau^3$, which occurred randomly):
\begin{align*}
\tau^1,\tau^3=&(s_{BAD},try), (s_{BAD},try), (s_{BAD},try),\\
&(s_{BAD},try), (s_{BAD},try), (s_{BAD},try),
(s_{BAD},try),\\ &(s_{BAD},try), (s_{BAD},try),
 (s_{BAD},try), s_{BAD}
\end{align*}
\begin{align*}
\tau^2=&(s_{BAD},try), (s_{GOOD},try), (s_{BAD},try),
(s_{BAD},try),\\ &(s_{BAD},try), (s_{BAD},try),
(s_{BAD},try), (s_{BAD},try),\\ &(s_{BAD},try),
(s_{BAD},try), s_{BAD}
\end{align*}
Tables \ref{tbl:SC-SB} and \ref{tbl:SC-AB} show all solutions that were Pareto efficient in at least two runs, for $\mathrm{Obj}^S$ and $\mathrm{Obj}^A$ respectively.  The results emphasize the distinction between the two objective functions.  In Table \ref{tbl:SC-SB} the correct formula $\always~good$ is Pareto efficient in two runs, but in most runs the obviously-incorrect $\always~\bot$ is the only Pareto efficient formula (and note that $\mathrm{Obj}^S(\always~\bot) \approxeq\mathrm{Obj}^S(\always~good)$).  In contrast, Table \ref{tbl:SC-AB} shows that when using $\mathrm{Obj}^A$, the true function $\always~good$ is Pareto efficient in all twenty runs.
\begin{table}
	\begin{center}
		\caption{Pareto efficient solutions in state-based SlimChance}	\label{tbl:SC-SB}
		\begin{tabular}{c|c|c|c}
			Formula $\phi$ & $\mathrm{Obj}^S(\phi)$ & $FC(\phi)$ & \# Runs \\
			\hline \hline
			$\always \bot$ & -0.3139852 & 2 & 18\\
			$\always~good$ & -0.3139852 & 2 & 2
		\end{tabular}
	\end{center}
\end{table}
\begin{table}
	\begin{center}
		\caption{Pareto efficient solutions in action-based SlimChance}	\label{tbl:SC-AB}
		\begin{tabular}{c|c|c|c}
			Formula $\phi$ & $\mathrm{Obj}^A(\phi)$ & $FC(\phi)$ & \# Runs \\
			\hline \hline
			$\always~good$ & -0.4623490 & 2 & 20 \\
			$\always(good \until (\Next~good))$ & -0.4939355 & 5 & 5 \\
			$\always(good \vee \Next~good)$ & -0.9400473 & 5 & 5 \\
			$\always((\Next~good)\until good)$ & -0.9400473 & 5 & 3 \\
			$\always((\Next~good) \vee good)$ & -0.9400424 & 5 & 2
		\end{tabular}
	\end{center}
\end{table}
\subsection{CleaningWorld domain}

In the CleaningWorld domain, the agent is a vacuum cleaning robot in a dirty room.  The room is characterized by some initial amount $dirt \in \mathbb{N}_0$ of dirt; the agent has some battery level $battery \in \mathbb{N}_0$.  The actions available to the agent are: $vacuum$, which reduces both $dirt$ and $battery$ by one; $dock$, which plugs the robot into a charger, allowing it to increment $battery$ for each time step it remains docked; $undock$, which unplugs the robot from the charger; $wait$, which allows the robot to remain docked if it is currently docked, but otherwise simply decrements $battery$.  If the robot's battery dies ($battery = 0$), the robot may only perform the dummy action $beDead$.  The domain has two propositions $batteryDead$, which is true iff $battery = 0$, and $roomClean$, which is true iff $dirt = 0$.  There are also propositions corresponding to each action (where, e.g., the proposition $vacuum$ is true whenever the agent's last action was to vacuum).  The agent is to satisfy the LTL objective $\always~((\Next~vacuum) \until roomClean)$.

An agent attempting to minimize violation cost for this specification produced three demonstration trajectories of 10 time steps each.  Because CleaningWorld is deterministic, all three trajectories were identical.  Here we represent each state $s$ by $(d,b)$ where $d$ is the amount of dirt still in the room in state $s$ and $b$ is the robot's current battery level.
\begin{align*}
\tau^1,\tau^2,\tau^3=&((5,3),vacuum), ((4,2),vacuum),\\ &((3,1),dock), ((3,1),wait), ((3,3),undock),\\  &((3,3),vacuum), ((2,2),vacuum), ((1,1),dock),\\ &((1,1),wait), ((1,3),undock), (1,3)\\
\end{align*}
\begin{table}
	\begin{center}
		\caption{Pareto efficient solutions in state-based CleaningWorld}	\label{tbl:CW-SB}
		\begin{tabular}{c|c|c|c}
			Formula $\phi$ & $\mathrm{Obj}^S(\phi)$ & $FC(\phi)$ & \# Runs \\
			\hline \hline
			$\always~roomClean$ & -208.69876 & 2 & 20 \\
			$\always(\eventually~roomClean)$ & -216.91139 & 3 & 20 \\
			$\always((\Next~roomClean) \vee vacuum)$ & -217.40816 & 5 & 2 \\
			$\always((\always~\top) \until roomClean)$ & -216.91169 & 5 & 2\\
			$\always(\eventually(undock \until roomClean))$ & -216.91170 & 5 & 2
		\end{tabular}
	\end{center}
	\vspace{-1em}
\end{table}
\begin{table}
	\begin{center}
		\caption{Pareto efficient solutions in action-based CleaningWorld}	\label{tbl:CW-AB}
		\begin{tabular}{c|c|c|c}
			Formula $\phi$ & $\mathrm{Obj}^A(\phi)$ & $FC(\phi)$ & \# Runs \\
			\hline \hline
			$\always(roomClean)$ & -72.74240 & 2 & 20 \\
			$\always(\eventually~roomClean)$ & -75.15686 & 3 & 20 \\
			$\always(vacuum \vee \eventually~roomClean)$ & -75.15832 & 5 & 3 \\
			$\always(\eventually (roomClean \vee dock))$ & -75.15782 & 5 & 3 \\
			$\always((\eventually~roomClean) \vee dock)$ & -75.15832 & 5 & 2 \\
			$\always((\Next roomClean) \vee vacuum)$ & -75.64639 & 5 & 2
		\end{tabular}
	\end{center}
	\vspace{-2em}
\end{table}
Tables \ref{tbl:CW-SB} and \ref{tbl:CW-AB} show all solutions that were Pareto efficient in at least two runs, for $\mathrm{Obj}^S$ and $\mathrm{Obj}^A$ respectively.  The formulas $\always~roomClean$ and $\always(\eventually~roomClean)$ are generated in all 20 runs by both $\mathrm{Obj}^S$ and $\mathrm{Obj}^A$.  These formulas (in particular, $\always~ roomClean$) arguably better describe agent behavior than the ``actual'' specification $\phi_{act}=\always((\Next~vacuum) \until roomClean)$: they are simpler than $\phi_{act}$ while generating identical trajectories.  This is reflected by the fact that $\phi_{act}$ \textit{was} generated by the algorithm for both state- and action-based runs, but $\mathrm{Obj}^S(\phi_{act})=-215.78773$, $\mathrm{Obj}^A(\phi_{act})=-75.10621$, and $FC(\phi_{act})=5$, which is Pareto dominated by $\always(\eventually~roomClean)$ when considering either $\mathrm{Obj}^S$ or $\mathrm{Obj}^A$. Perhaps because of this, the actual formula is never recovered (although similar formulas occasionally are, such as $\always((\Next roomClean) \vee vacuum)$).
\section{Discussion}

While for demonstration purposes we chose to use NSGA-II for optimization, in principle any algorithm that can optimize over LTL formulas should suffice.  Exploring other algorithms is a topic for future work.  In particular, the genetic programming methods employed operate entirely on the syntax of LTL; a method that can make some use of LTL semantics may find optimal solutions more efficiently.

Optimizing over the space of all LTL formulas is difficult because of the combinatorial nature of this space.  Since the number of LTL formulas of length $\ell$ increases exponentially in $\ell$, optimization algorithms like NSGA-II are likely to recover simple formulas that explain the demonstrator's behavior reasonably well, but are less likely to recover complex formulas that better explain the behavior.

We do not specify how to select between Pareto efficient solutions; this depends on the relative degree to which system designers value simplicity versus explanatory power.  In practice, system designers with clear preferences could convert the given problem into a single-objective problem with objective $f(\mathrm{Obj}(\phi), FC(\phi))$ where $f$ is some nondecreasing function encoding these preferences.

The major drawback of the proposed approach is its scalability.  Table \ref{tbl:runtime} indicates that evaluation on CleaningWorld with the action-based objective took, on average, roughly 9h 9m.  For problems with much larger state and action spaces, this approach is certainly intractable.  Theoretically, a single iteration in the computation of $\Viol^\pi_\phi$ takes time in $O(|S|^2|Q||U|)$.  Run time for objective function evaluation also scales linearly in the total number of demonstration time steps.  Identifying approaches with better theoretical and practical run times is a topic for future work.

This paper also assumes that the demonstrator is operating in an environment with complete information (e.g., an MDP), no other agents, and known transition dynamics.  Extensions to unknown transition dynamics, POMDPs, and multi-agent domains are a topic for future work.

In both given examples, the ``true'' specification can be modeled using a reward function: in SlimChance, give high reward if and only if the agent is in $s_{GOOD}$; in CleaningWorld, give high reward only when $roomClean$ is true.  IRL may easily recover these reward functions, and would likely converge more quickly than our approach.  These examples are meant more to show the viability of the proposed approach than its superiority to IRL in these domains.

While the given problem assumes that the apprentice passively observes the demonstrator's trajectories, future work could consider an active learning approach, in which the apprentice (for example) poses new MDPs involving the same predicates (or perturbs the given MDP), and `asks' the demonstrator to generate trajectories in the posed MDPs.

\section{Conclusion}

In this paper, we introduced the problem of inferring linear temporal logic (LTL) specifications from agent behavior in Markov Decision Processes as a road to interpretable apprenticeship learning, combining the representational power and interpretability of temporal logic with the generalizability of inverse reinforcement learning.  We formulated this as a two-objective optimization problem, and introduced objective functions using a notion of ``violation cost'' to quantify the ability of an LTL formula to explain demonstrated behavior.  We presented results using genetic programming to solve this problem in the SlimChance and CleaningWorld domains.

\section{Acknowledgements}
This project was in part supported by ONR grant N00014-16-1-2278.

\bibliography{cdc2017}

\begin{thebibliography}{10}

\bibitem{Abbeel2004}
Pieter Abbeel and Andrew~Y Ng.
\newblock {Apprenticeship learning via inverse reinforcement learning}.
\newblock {\em Proc. 21st International Conference on Machine Learning (ICML)},
  pages 1--8, 2004.

\bibitem{Arnold2017}
Thomas Arnold, Daniel Kasenberg, and Matthias Scheutz.
\newblock Value alignment or misalignment--what will keep systems accountable?
\newblock In {\em 3rd International Workshop on AI, Ethics, and Society}, 2017.

\bibitem{Baier2008}
Christel Baier and Joost-Pieter Katoen.
\newblock {\em Principles of Model Checking}.
\newblock The MIT Press, 2008.

\bibitem{gecco2015}
Daniil Chivilikhin, Ilya Ivanov, and Anatoly Shalyto.
\newblock {Inferring Temporal Properties of Finite-State Machine Models with
  Genetic Programming}.
\newblock In {\em Proc. 2015 Annual Conference on Genetic and Evolutionary
  Computation}, pages 1185--1188, 2015.

\bibitem{Ding2011}
Xu~Chu Ding, Stephen~L. Smith, Calin Belta, and Daniela Rus.
\newblock {LTL control in uncertain environments with probabilistic
  satisfaction guarantees}.
\newblock In {\em Proceedings - IFAC World Congress}, volume~18, pages
  3515--3520, 2011.

\bibitem{Dzifcak2009}
Juraj Dzifcak, Matthias Scheutz, Chitta Baral, and Paul Schermerhorn.
\newblock {What to do and how to do it: Translating natural language directives
  into temporal and dynamic logic representation for goal management and action
  execution}.
\newblock In {\em Proceedings - IEEE International Conference on Robotics and
  Automation}, pages 4163--4168, 2009.

\bibitem{rabinizer3}
Javier Esparza and Jan Křet{\'{i}}nsk{\'{y}}.
\newblock {From LTL to deterministic automata: A safraless compositional
  approach}.
\newblock In {\em Lecture Notes in Computer Science (including subseries
  Lecture Notes in Artificial Intelligence and Lecture Notes in
  Bioinformatics)}, volume 8559, pages 192--208. Springer International
  Publishing, 2014.

\bibitem{Fu2014}
Jie Fu and Ufuk Topcu.
\newblock {Probably Approximately Correct MDP Learning and Control With
  Temporal Logic Constraints}.
\newblock In {\em Robotics: Science and Systems X}, 2014.

\bibitem{Gabel2008Symbolic}
Mark Gabel and Zhendong Su.
\newblock Symbolic mining of temporal specifications.
\newblock In {\em Proc. 30th International Conference on Software Engineering},
  ICSE '08, pages 51--60, New York, NY, USA, 2008. ACM.

\bibitem{Gabel2010}
Mark Gabel and Zhendong Su.
\newblock Online inference and enforcement of temporal properties.
\newblock In {\em Proceedings of the 32Nd ACM/IEEE International Conference on
  Software Engineering - Volume 1}, ICSE '10, pages 15--24, New York, NY, USA,
  2010. ACM.

\bibitem{Guo2014}
M.~Guo and D.~V. Dimarogonas.
\newblock {Multi-agent plan reconfiguration under local LTL specifications}.
\newblock {\em The International Journal of Robotics Research}, 34(2):218--235,
  2014.

\bibitem{hadka2012}
David Hadka.
\newblock Moea framework: a free and open source java framework for
  multiobjective optimization, 2012.

\bibitem{Kong2014}
Zhaodan Kong, Austin Jones, Ana {Medina Ayala}, Ebru {Aydin Gol}, and Calin
  Belta.
\newblock {Temporal Logic Inference for Classification and Prediction from
  Data}.
\newblock {\em Proceedings of the 17th International Conference on Hybrid
  Systems: Computation and Control}, pages 273--282, 2014.

\bibitem{koza1992}
John~R Koza.
\newblock {\em Genetic programming: on the programming of computers by means of
  natural selection}, volume~1.
\newblock MIT press, 1992.

\bibitem{Lahijanian2015}
Morteza Lahijanian, Shaull Almagor, Dror Fried, Lydia~E Kavraki, and Moshe~Y
  Vardi.
\newblock {This Time the Robot Settles for a Cost: A Quantitative Approach to
  Temporal Logic Planning with Partial Satisfaction}.
\newblock In {\em Proceedings of the AAAI Conference on Artificial
  Intelligence}, volume~29, pages 3664--3671, 2015.

\bibitem{Leahy2015}
Kevin Leahy, Austin Jones, Mac Schwager, and Calin Belta.
\newblock {Distributed Information Gathering Policies under Temporal Logic
  Constraints}.
\newblock In {\em IEEE Conference on Decision and Control (CDC)}, volume~54,
  pages 6803--6808, 2015.

\bibitem{lemieux2015}
Caroline Lemieux, Dennis Park, and Ivan Beschastnikh.
\newblock General ltl specification mining.
\newblock In {\em Automated Software Engineering (ASE), 30th IEEE/ACM
  International Conference on}, pages 81--92. IEEE, 2015.

\bibitem{burlap}
James MacGlashan.
\newblock {Brown-UMBC Reinforcement Learning and Planning (BURLAP)}, 2016.

\bibitem{Ng2000}
Andrew Ng and Stuart Russell.
\newblock {Algorithms for inverse reinforcement learning}.
\newblock In {\em Proc. Seventeenth International Conference on Machine
  Learning}, volume~0, pages 663--670, 2000.

\bibitem{LTL}
Amir Pnueli.
\newblock {The temporal logic of programs}.
\newblock In {\em 18th Annual Symposium on Foundations of Computer Science},
  pages 46--57, 1977.

\bibitem{ReyesCastro2013}
Luis~I. {Reyes Castro}, Pratik Chaudhari, Jana T{\"{u}}mov{\'{a}}, Sertac
  Karaman, Emilio Frazzoli, and Daniela Rus.
\newblock {Incremental sampling-based algorithm for minimum-violation motion
  planning}.
\newblock In {\em Proc. IEEE Conference on Decision and Control}, pages
  3217--3224, 2013.

\bibitem{Sharan2014}
Rangoli Sharan and Joel Burdick.
\newblock {Finite state control of POMDPs with LTL specifications}.
\newblock In {\em Proceedings of the American Control Conference}, pages
  501--508, 2014.

\bibitem{Svorenova2015}
M{\'{a}}ria Svore{\v{n}}ov{\'{a}}, Martin Chmel{\'{i}}k, Kevin Leahy,
  Hasan~Ferit Eniser, Krishnendu Chatterjee, Ivana {\v{C}}ern{\'{a}}, and Calin
  Belta.
\newblock {Temporal logic motion planning using POMDPs with parity objectives}.
\newblock In {\em Proceedings of the 18th International Conference on Hybrid
  Systems Computation and Control}, pages 233--238, 2015.

\bibitem{Tumova2013}
Jana Tumova, Gavin~C Hall, Sertac Karaman, Emilio Frazzoli, and Daniela Rus.
\newblock {Least-violating control strategy synthesis with safety rules}.
\newblock In {\em Proceedings of the 16th International Conference on Hybrid
  Systems: Computation and Control}, pages 1--10, 2013.

\bibitem{Wolff2012}
Eric~M. Wolff, Ufuk Topcu, and Richard~M. Murray.
\newblock {Robust control of uncertain Markov Decision Processes with temporal
  logic specifications}.
\newblock In {\em IEEE Conference on Decision and Control (CDC)}, volume~51,
  pages 3372--3379, 2012.

\end{thebibliography}
\bibliographystyle{plain}
\end{document}